\documentclass[12pt]{article}   
\usepackage[utf8]{inputenc}
\usepackage[T1]{fontenc}
\usepackage{graphicx}
\usepackage{epstopdf} 
\usepackage{color}
\input{epsf}
\usepackage{lmodern}
\usepackage{amsmath}
\newcommand{\be}{\begin{equation}}
\newcommand{\ee}{\end{equation}}
\newcommand{\bear}{\begin{eqnarray}}
\newcommand{\ear}{\end{eqnarray}}

\begin{document}
\title{Generalization of the least uncomfortable journey problem}
\author{Nivaldo A. Lemos  \\
\small
{\it Instituto de F\'{\i}sica - Universidade Federal Fluminense}\\
\small
{\it Av. Litor\^anea, S/N, Boa Viagem, 
24210-340, Niter\'oi - Rio de Janeiro, Brasil}\\
\small
{\it  nivaldolemos@id.uff.br}}

\date{\today}

\maketitle

\begin{abstract}

The variational problem of the least uncomfortable journey between two locations on a straight line is simplified by a choice of the dependent variable. It is shown that taking the position, instead of the velocity, as the optimal function of time to be determined does away with the isoperimetric constraint.  The same results as those found with the velocity as the dependent variable are obtained in a simpler and more concise way. Next the problem is generalized for motion on an arbitrary curve. In the case of  acceleration-induced discomfort, it is shown that, as expected, motion on a curved path is always more uncomfortable than motion on a  straight line. 
It is not clear that this is  necessarily the case for jerk-induced discomfort, which appears to indicate that the acceleration provides a more reasonable measure of  the discomfort than the jerk. The example of motion on a circular path is studied. Although we have been unable to solve the problem analytically, approximate solutions have been constructed by means of trial functions and the exact solution has been found numerically for some choices of the relevant parameters.

\end{abstract}

\maketitle

\section{Introduction}\label{Intro}

The calculus of variations is a powerful mathematical tool that finds application in virtually every branch  of theoretical physics. 
A variational problem of recent invention is that of  finding  the least uncomfortable way to travel from point $A$ to point $B$ on a straight line, with both the travel time and the distance between the two points 
fixed \cite{Anderson}.  


Since frequent acceleration and deceleration  make a trip
uncomfortable, the discomfort can be  quantified by the integral of the square of the acceleration taken over the duration of the journey. 
Abrupt changes of a constant acceleration are  also a source of distress. Therefore, another possible quantification of the discomfort is in terms of the time rate of change of the acceleration, usually known as jerk. In this  case the discomfort is defined by  the square of the jerk integrated over the duration of the journey.  


The choice of independent variable impacts a variational problem. Sometimes an unfortunate choice makes the problem seem much more difficult than it actually is.
In \cite{Lemos1} it is shown that with the time as independent variable 
the least uncomfortable journey problem becomes much simpler than with  the position as independent variable, which was the original choice of Anderson, Desaix and Nyqvist \cite{Anderson}.

In both  \cite{Anderson} and  \cite{Lemos1} the problem of the least uncomfortable journey was understood as a search for the optimal velocity, which minimizes the discomfort. Either with position \cite{Anderson} or time \cite{Lemos1} as independent variable, this approach is characterized by the existence of an isoperimetric constraint, which is taken into account by the method of Lagrange multipliers. It turns out that the choice of dependent variable, namely the unknown function to be determined, also impacts a variational problem. We show that with time as independent variable and position as dependent variable, 
the isoperimetric constraint is automatically satisfied and, both for acceleration-induced and jerk-induced discomfort, one has to deal with an ordinary unconstrained variational problem. In the case in which  the discomfort is measured by the jerk, natural boundary conditions arise because the acceleration cannot be prescribed either at the beginning or the end of the journey. For both choices of the discomfort functional the same results as those found in \cite{Lemos1} are obtained in a more straighforward way, without having to bother about constraints. 

Next we generalize the problem by considering motion on an arbitrary curve with the arc length $s$ as the dependent variable. Both for discomfort measured by the acceleration and by the jerk, it so happens that if the curvature is not constant  the velocity $v = {\dot s}$ cannot be taken as the dependent variable. In the case of acceleration-induced discomfort, it is found that, as expected, motion on a curved path gives rise to more  discomfort than motion on a straight line.  It is somewhat surprising that this is not necessarily the case for jerk-induced discomfort. This leads us to believe that  it is the acceleration, rather than the jerk, that provides the more trustworthy measure of the discomfort. As a case study, we consider motion on a circular path. In spite of some advances towards the exact solution, we have been unable to express it in an explicit and manageable form. Therefore,  approximate solutions  have been constructed by means of trial functions and compared with the exact solutions obtained numerically. 

The paper is organized as follows. In Section \ref{Original} we recall the statement of the problem for motion on a straight line as the search for the optimal velocity function  $v(t)$ that minimizes the discomfort. In Section \ref{Position-dep-var} we rephrase the problem as the search for the optimal position function $x(t)$ that leads to the least discomfort, and it is pointed out that the isoperimetric constraint is automatically taken care of.  The exact solution is found both for the acceleration-induced and  jerk-induced  discomfort, in the latter case taking into account  the natural boundary conditions, which are derived. In Section \ref{Generalization} the least uncomfortable journey problem is generalized for motion on an arbitrary curve.  
 With the arc length $s$
as the dependent variable and time as the independent variable, it is shown that the acceleration-induced discomfort depends only on the curvature and is always larger for a curved trajectory than for a straight line. The jerk-induced discomfort  depends both on  the curvature and the torsion, but it is not always necessarily larger for a curved path than for a straight line. In Section \ref{Circular} the case of motion on a circular path is studied 
and  approximate trial-function solutions are given for the optimal function $s(t)$   which are compared with the exact numerical solutions.
Section \ref{Final-remarks} is dedicated to a few final remarks.

\section{Original statement of the   least uncomfortable journey problem}\label{Original}
 
The problem of  finding the least uncomfortable way to travel between two points on a straight line  was posed by Anderson, Desaix, and Nyqvist  \cite{Anderson}. On a straight road, a vehicle departs from point $A$ at $t=0$ and  arrives at point $B$
 when $t=T$. Let the coordinate system be so chosen that the departure point $A$ corresponds to  $x= 0$ and the arrival point $B$ corresponds 
to  $x= D$.
The travel time $T$ is  fixed. The problem consists in finding the velocity $v(t)$  that minimizes  the discomfort functional defined by
\begin{equation}
\label{discomfort-functional}
A[v] = \int_0^T {\dot v}^2 dt 
\end{equation}	
with the boundary conditions
\begin{equation}
\label{boundary-conditions-acceleration}
v(0) = 0, \,\,\,\,\,\, v(T) = 0, 
\end{equation}
and under the isoperimetric constraint
\begin{equation}
\label{isoperimetric-constraint-acceleration}
\int_0^T v dt = D. 
\end{equation}

If the discomfort is measured by the integral of the square of the jerk, the problem is the same as above except that now the discomfort functional to be minimized is
\begin{equation}
\label{discomfort-functional-jerk}
J[v] = \int_0^T {\ddot v}^2 dt 
\end{equation}	
 under the same isoperimetric constraint (\ref{isoperimetric-constraint-acceleration})
and, as argued in \cite{Lemos1},
the boundary conditions
\begin{equation}
\label{boundary-conditions-functional-jerk}
v(0) = 0, \,\,\,\,\,\, v(T) = 0, \,\,\,\,\,\, {\ddot v}(0) = 0, \,\,\,\,\,\, {\ddot v}(T) = 0.
\end{equation}

In both cases the unknown function is the velocity $v$. With this choice of dependent variable, an essential feature of the problem is the isoperimetric constraint (\ref{isoperimetric-constraint-acceleration}). 

The  treatment of the problem in \cite{Anderson} was based on choosing the position $x$ as the independent variable and searching for the optimal velocity $v(x)$ that minimizes the discomfort. It is a fact that sometimes a judicious choice of independent variable makes a variational problem more tractable.  In \cite{Lemos1} it is shown that choosing the time as the independent variable  and searching for the optimal velocity $v(t)$
leads to a drastic simplification of the problem, especially in the case in which the  discomfort is measured by the jerk. Similarly,  the choice of dependent variable, that is, the function to be determined, may simplify a variational problem. As we proceed to show, if one looks for the position as a function of time the isoperimetric constraint disappears and one is led  to a simpler unconstrained higher-derivative variational problem.

\section{Position as the dependent variable}\label{Position-dep-var}

Let us take the time $t$ as the independent variable, the position $x$ as the dependent variable and  search for the function  $x(t)$  that yields the minimum discomfort.

\subsection{Discomfort measured by the acceleration}

 The  functional to be minimized is
\begin{equation}
\label{discomfort-functional-acceleration-x}
A[x] = \int_0^T {a(t)}^2 dt =  \int_0^T {{\ddot x}(t)}^2 dt
\end{equation}	
with the boundary conditions
\begin{equation}
\label{boundary-conditions-acceleration-x}
x(0) = 0, \,\,\,\,\,\, x(T) = D. 
\end{equation}
Since the journey starts from rest and ends at rest, we have the additional boundary conditions
\begin{equation}
\label{boundary-conditions-acceleration-x-dot}
{\dot x}(0) = 0, \,\,\,\,\,\, {\dot x}(T) = 0. 
\end{equation}
Note that by taking $x$ as the dependent variable the isoperimetric constraint (\ref{isoperimetric-constraint-acceleration}) is automatically incorporated in the boundary conditions (\ref{boundary-conditions-acceleration-x}). This means that now we are dealing with a common unconstrained higher-derivative variational problem, since the functional to be minimized depends on the second derivative of the unknown function.  

For a functional of the form
\begin{equation}
\label{functional-second-order}
A[x] = \int_0^T L(x, {\dot x}, {\ddot x}, t) dt 
\end{equation}
under the standard boundary conditions of fixed values for $x(t)$ and ${\dot x}(t)$ at $t=0$ and $t=T$, it is well known \cite[Section 4.1]{Kot}  that a necessary condition for a function $x(t)$ to minimize the functional (\ref{functional-second-order})  is that it obey the generalized Euler-Lagrange equation
\begin{equation}
\label{Euler-Lagrange-second-order}
\frac{d^2}{dt^2}\bigg(\frac{\partial L}{\partial {\ddot x}} \bigg) - 
\frac{d}{dt}\bigg(\frac{\partial L}{\partial {\dot x}} \bigg) + \frac{\partial L}{\partial x} =0.
\end{equation}

For the discomfort functional (\ref{discomfort-functional-acceleration-x}) we have
\begin{equation}
\label{Lgrangian-second-order}
L = {\ddot x}^2, 
\end{equation}
and equation (\ref{Euler-Lagrange-second-order}) becomes simply
\begin{equation}
\label{Euler-Lagrange-second-order-acceleration}
\frac{d^4x}{dt^4} =0.
\end{equation}
The general solution to this equation is
\begin{equation}
\label{Euler-Lagrange-second-order-acceleration-general-solution}
x(t)  = a_0 +a_1t+a_2t^2+a_3t^3,
\end{equation}
where $a_0, a_1, a_2, a_3$ are arbitrary constants. Application of the boundary conditions (\ref{boundary-conditions-acceleration-x}) and (\ref{boundary-conditions-acceleration-x-dot}) leads easily to
\begin{equation}
\label{x-of-t-least-uncomfortable-x}
x (t) = D \bigg(   3\frac{t^2}{T^2} - 2\frac{t^3}{T^3}\bigg),
\end{equation}
which is the same result found in \cite{Lemos1}, in which the original treatment with $v$ as the dependent variable was adopted.
The minimum value of the discomfort is obtained by inserting the above $x(t)$ into the right-hand side of equation (\ref{discomfort-functional-acceleration-x}). An easy calculation yields
\begin{equation}
\label{minimum-A-discomfort}
A_{min} = 12 \frac{D^2}{T^3}.
\end{equation}  

\subsection{Discomfort measured by the jerk}

Now the  functional to be minimized is
\begin{equation}
\label{discomfort-functional-jerk-x}
J[x] = \int_0^T {{\dot a}(t)}^2 dt =  \int_0^T {{\dddot x}(t)}^2 dt
\end{equation}
with the boundary conditions (\ref{boundary-conditions-acceleration-x}) and (\ref{boundary-conditions-acceleration-x-dot}).

For a functional of the form
\begin{equation}
\label{functional-third-order}
J[x] = \int_0^T L(x, {\dot x}, {\ddot x}, {\dddot x}, t) dt,
\end{equation}
the minimizer function $x(t)$ has to satisfy the  generalized Euler-Lagrange equation
\begin{equation}
\label{Euler-Lagrange-third-order}
-\frac{d^3}{dt^3}\bigg(\frac{\partial L}{\partial {\dddot x}} \bigg) + \frac{d^2}{dt^2}\bigg(\frac{\partial L}{\partial {\ddot x}} \bigg) - 
\frac{d}{dt}\bigg(\frac{\partial L}{\partial {\dot x}} \bigg) + \frac{\partial L}{\partial x} =0.
\end{equation}
This is a sixth-order ordinary differential equation, and the boundary conditions (\ref{boundary-conditions-acceleration-x}) and (\ref{boundary-conditions-acceleration-x-dot}) are insufficient to determine a unique solution for $x(t)$. The additional boundary conditions would seem to be \cite{Anderson}
\begin{equation}
\label{boundary-conditions-ADN}
{\ddot x}(0)  = 0, \,\,\,\,\,\, {\ddot x}(T) = 0.
\end{equation}
 These, together with (\ref{boundary-conditions-acceleration-x}) and (\ref{boundary-conditions-acceleration-x-dot}),  would comprise the standard boundary conditions for a variational problem involving the third derivative of the unknown function. However, as argued in \cite{Lemos1},   the specification of the initial acceleration is at odds with the basic principles of Newtonian mechanics, and its enforcement might require  pathological, unphysical forces.

The right thing to do \cite{Lemos1} is leave ${\ddot x}(0)$ and ${\ddot x}(T)$ free, which leads to a variational problem with variable endpoints as concerns the values of ${\ddot x}(t)$ at $t=0$ and $t=T$. Variable endpoints give rise to  what are commonly called natural boundary conditions \cite[Section 9.1]{Kot}.

Let us present a short derivation of the natural boundary conditions for the functional (\ref{functional-third-order}) with ${\ddot x}(0)$ and ${\ddot x}(T)$ free. Because of the boundary conditions (\ref{boundary-conditions-acceleration-x}) and (\ref{boundary-conditions-acceleration-x-dot}) the variation $\delta x$ is required to satisfy
\begin{equation}
\label{boundary-conditions_variations-jerk}
{\delta x}(0)  = 0, \,\,\,\,\,\, {\delta x}(T) = 0,  \,\,\,\,\,\, {\delta {\dot x}}(0) = 0, \,\,\,\,\,\, {\delta {\dot x}}(T) = 0.
\end{equation}
The variation of the functional (\ref{functional-third-order}) is
\begin{equation}
\label{variation-functional-third-derivative}
\delta J = \int_0^T \bigg( \frac{\partial L}{\partial x} \delta x + 
\frac{\partial L}{\partial {\dot x}} {\delta {\dot x}} + 
+ \frac{\partial L}{\partial {\ddot x}} {\delta {\ddot x}} 
+ \frac{\partial L}{\partial {\dddot x}} {\delta {\dddot x}}
\bigg)  dt.
\end{equation}
Successive integrations by parts lead to
\begin{eqnarray}
\label{variation-functional-third-derivative-integrated-by-parts}
\delta J & = & \frac{\partial L}{\partial {\dddot x}} {\delta {\ddot x}}\bigg\vert_0^T
+ \bigg(\frac{\partial L}{\partial {\ddot x}} - \frac{d}{dt}\frac{\partial L}{\partial {\dddot x}} \bigg) \delta {\dot x}\bigg\vert_0^T + 
\bigg(\frac{\partial L}{\partial {\dot x}} - \frac{d}{dt}\frac{\partial L}{\partial {\ddot x}} + \frac{d^2}{dt^2}\frac{\partial L}{\partial {\dddot x}}\bigg) \delta x\bigg\vert_0^T    \nonumber \\
 & & + \int_0^T \bigg[ -\frac{d^3}{dt^3}\bigg(\frac{\partial L}{\partial {\dddot x}} \bigg) + \frac{d^2}{dt^2}\bigg(\frac{\partial L}{\partial {\ddot x}} \bigg) - 
\frac{d}{dt}\bigg(\frac{\partial L}{\partial {\dot x}} \bigg) + \frac{\partial L}{\partial x} \bigg] \delta x dt.
\end{eqnarray}
With the use of (\ref{boundary-conditions_variations-jerk}) this reduces to
\begin{equation}
\label{variation-functional-third-derivative-integrated-by-parts-reduced}
\delta J = \frac{\partial L}{\partial {\dddot x}} {\delta {\ddot x}}\bigg\vert_0^T
+  \int_0^T \bigg[ -\frac{d^3}{dt^3}\bigg(\frac{\partial L}{\partial {\dddot x}} \bigg) + \frac{d^2}{dt^2}\bigg(\frac{\partial L}{\partial {\ddot x}} \bigg) - 
\frac{d}{dt}\bigg(\frac{\partial L}{\partial {\dot x}} \bigg) + \frac{\partial L}{\partial x} \bigg] \delta x dt.
\end{equation}
If $x(t)$ is to minimize $J$ one must have $\delta J = 0$ for arbitrary $\delta x$ with arbitrary values of $\delta {\ddot x}(t)$ at $t=0$ and $t=T$. This means that we are free to choose the variation $\delta x$ such that $\delta {\ddot x} =0$ at both $t=0$ and $t=T$. In this case the boundary term on the right-hand side of  (\ref{variation-functional-third-derivative-integrated-by-parts-reduced})
vanishes. Apart from the conditions at $t=0$ and $t=T$, the variation $\delta x $ is an otherwise arbitrary function. Then   the fundamental lemma of the calculus of variations \cite[p. 39]{Kot} establishes that  the  condition $\delta J = 0$, which is necessary for a minimum, implies that the coefficient of $\delta x$ in the integral on the right-hand side of (\ref{variation-functional-third-derivative-integrated-by-parts-reduced}) vanishes. Consequently, the minimizer $x(t)$ must satisfy the previously announced  generalized Euler-Lagrange equation (\ref{Euler-Lagrange-third-order}). This, in turn, further reduces $\delta J$ to 
\begin{equation}
\label{variation-functional-third-derivative-integrated-by-parts-reduced-further}
\delta J = \frac{\partial L}{\partial {\dddot x}} {\delta {\ddot x}}\bigg\vert_0^T.
\end{equation}
Since the values of $\delta {\ddot x} (t)$ are arbitrary at $t=0$ and $t=T$, we can choose the variation $\delta x(t)$ in such a way that $\delta{\ddot x} (0) \neq 0, \delta {\ddot x} (T) = 0$ and vice versa. Therefore, the  condition  $\delta J = 0$ requires that the following boundary conditions be obeyed:
\begin{equation}
\label{variation-functional-higher-derivative-natural-boundary-x}
\frac{\partial L}{\partial {\dddot x}}\bigg\vert_{t=0} = 0, \,\,\,\,\, \frac{\partial L}{\partial {\dddot x}}\bigg\vert_{t=T} = 0.
\end{equation}
These are {\it natural boundary conditions} \cite[p. 194]{Kot} for the problem with
variable endpoints as regards ${\ddot x}(t)$. 

For the discomfort functional (\ref{discomfort-functional-jerk-x}) we have
\begin{equation}
\label{Lagrangian-third-order-derivative}
L=  {\dddot x}^2
\end{equation}
and the natural boundary conditions (\ref{variation-functional-higher-derivative-natural-boundary-x}) become
\begin{equation}
\label{natural-boundary-conditions-jerk-x}
{\dddot x}(0)  = 0, \,\,\,\,\,\, {\dddot x}(T) = 0.
\end{equation}
These are exactly the same natural boundary conditions as those expressed by equation (24) in \cite{Lemos1}, where they were derived on the basis of taking the velocity $v$ as the dependent variable.

With $L$ given by (\ref{Lagrangian-third-order-derivative}) it follows that the generalized Euler-Lagrange equation (\ref{Euler-Lagrange-third-order}) yields
\begin{equation}
\label{differential-equation-x}
\frac{d^6x}{dt^6}  = 0.
\end{equation}
The general solution to this equation is 
\begin{equation}
\label{Euler-Lagrange-third-order-acceleration-general-solution}
x(t)  = a_0 +a_1t+a_2t^2+a_3t^3+a_4t^4+a_5t^5.
\end{equation}
The six initially arbitrary constants are readily found by imposing the boundary conditions (\ref{boundary-conditions-acceleration-x}),  (\ref{boundary-conditions-acceleration-x-dot}) and (\ref{natural-boundary-conditions-jerk-x}). After a little algebra the result is
\begin{equation}
\label{x-of-t-least-uncomfortable-jerk}
x = \frac{D}{2} \bigg( 5\frac{t^2}{T^2} - 5\frac{t^4}{T^4} + 2\frac{t^5}{T^5}\bigg),
\end{equation}
which coincides with equation (41) in \cite{Lemos1}.

\subsection{Comments}

We have just shown that by taking as dependent variable --- that is, the unknown function to be determined --- the position $x$ instead of the velocity $v$, the isoperimetric constraint is eliminated and, in the case of the discomfort measured by the acceleration,  one is left with an ordinary higher-derivative variational problem. Just as the choice of independent variable, a proper choice of dependent variable may simplify a variational problem and eliminate constraints by having them automatically satisfied. If the discomfort is measured by the jerk, taking $x$ as the dependent variable also eliminates the isoperimetric 
constraint and leads to a higher-derivative variational problem with  variable endpoints as concerns $\ddot x$. The natural boundary conditions in this case are exactly the same as those found by treating $v$ as the dependent variable.

For our treatment of the problem with the position as dependent variable, it is also straightforward to prove, as in \cite{Lemos2}, that the optimal positions (\ref{x-of-t-least-uncomfortable-x}) and (\ref{x-of-t-least-uncomfortable-jerk}) actually yield, in each case, the minimum discomfort.

\section{Generalization of the least uncomfortable journey problem}\label{Generalization}

Let us suppose the motion takes place along a prescribe  curve. It will be convenient to parameterize the curve by the arc length $s$:
\begin{equation}
\label{arc-length-parameterized-curve}
{\bf r}(s) = x(s) {\hat {\bf x}} + y(s) {\hat {\bf y}}  + z(s) {\hat {\bf z}}.
\end{equation}
The motion  is completely characterized once the function $s(t)$ is given.

\subsection{Discomfort induced by the acceleration}

For the velocity we have
\begin{equation}
\label{velocity-arc-length}
{\bf v} = \frac{d{\bf r}}{dt} = \frac{d \bf r}{ds}\frac{ds}{dt} = {\bf r}^{\prime}(s) {\dot s}.
\end{equation}
The acceleration is given by
\begin{equation}
\label{acceleration-arc-length}
{\bf a} = \frac{d{\bf v}}{dt} = {\bf r}^{\prime}(s) {\ddot s} + \frac{d {\bf r}^{\prime}(s)}{dt} {\dot s} = {\bf r}^{\prime}(s) {\ddot s} + {\bf r}^{\prime\prime}(s){\dot s}^2. 
\end{equation}

We shall first consider  the case in which the discomfort is measured by the acceleration, which means that the discomfort functional to be minimized is
\begin{equation}
\label{discomfort-functional-acceleration-generalized}
A[s] =\int_0^T \vert {\bf a} \vert^2 dt. 
\end{equation}
From (\ref{acceleration-arc-length}) it follows that 
\begin{equation}
\label{square-acceleration-arc-length}
\vert {\bf a} \vert^2 = {\bf a} \cdot {\bf a} = {\bf r}^{\prime}(s) \cdot {\bf r}^{\prime}(s) {\ddot s}^2 + 2 {\bf r}^{\prime}(s) \cdot {\bf r}^{\prime\prime}(s) {\dot s}^2{\ddot s} + {\bf r}^{\prime\prime}(s) \cdot {\bf r}^{\prime\prime}(s) {\dot s}^4
\end{equation}
Since
\begin{equation}
\label{unit-r-prime-s}
{\bf r}^{\prime}(s) \cdot {\bf r}^{\prime}(s) = \frac{d{\bf r}}{ds} \cdot \frac{d{\bf r}}{ds} = \frac{d{\bf r} \cdot d{\bf r}}{ds^2} = \frac{ds^2}{ds^2} =1,
\end{equation}
upon differentiation with respect to $s$ this equation yields
\begin{equation}
\label{orthogonality-r-prime-r-primeprime}
{\bf r}^{\prime}(s) \cdot {\bf r}^{\prime\prime}(s) = 0.
\end{equation}
Therefore,
\begin{equation}
\label{square-acceleration-arc-length-simplified}
\vert {\bf a} \vert^2 = {\ddot s}^2 +  {\bf r}^{\prime\prime}(s) \cdot {\bf r}^{\prime\prime}(s) {\dot s}^4.
\end{equation}
By definition 
\begin{equation}
\label{f(s)-defined}
 {\bf r}^{\prime\prime}(s) \cdot {\bf r}^{\prime\prime}(s) = {\kappa}^2(s),
\end{equation}
where $\kappa (s)$ is the curvature of the curve. Therefore, the discomfort functional (\ref{discomfort-functional-acceleration-generalized}) becomes 
\begin{equation}
\label{discomfort-functional-acceleration-generalized-arc-length}
A[s] =\int_0^T [ {\ddot s}^2 + {\kappa}^2(s) {\dot s} ^4 ] dt. 
\end{equation}

The boundary conditions are
\begin{equation}
\label{boundary-conditions-s}
 s(0)  = 0, \,\,\,\,\,\, s(T) = D
\end{equation}
as well as 
\begin{equation}
\label{boundary-conditions-s-dot}
{\dot s}(0)  = 0, \,\,\,\,\,\, {\dot s}(T) = 0
\end{equation}
inasmuch as the velocity is zero both at the beginning and at the end of the journey.

Equation (\ref{discomfort-functional-acceleration-generalized-arc-length}) shows that, as expected, curvature adds to the discomfort: if $\kappa \neq 0$ the discomfort given by (\ref{discomfort-functional-acceleration-generalized-arc-length}) is larger than the discomfort for motion on a straight line, which is given by (\ref{discomfort-functional-acceleration-x}) with $x=s$.

It is to be noted that, unless $\kappa (s)$ is a constant, as for a circle or a helix,  the discomfort
(\ref{discomfort-functional-acceleration-generalized-arc-length}) cannot be regarded as a functional of the speed $v={\dot s}$ alone, that is, the approach followed in \cite{Anderson} and \cite{Lemos1} is impossible.

It is a straightforward exercise to show that with
\begin{equation}
\label{Lagrangian-s}
L = {\ddot s}^2 + {\kappa}^2(s) {\dot s} ^4  
\end{equation}
the generalized Euler-Lagrange equation 
\begin{equation}
\label{Euler-Lagrange-s}
 \frac{d^2}{dt^2}\bigg(\frac{\partial L}{\partial {\ddot s}} \bigg) - 
\frac{d}{dt}\bigg(\frac{\partial L}{\partial {\dot s}} \bigg) + \frac{\partial L}{\partial s} =0
\end{equation}
 yields
\begin{equation}
\label{Euler-Lagrange-differential-equation-for-s}
 \frac{d^4 s}{dt^4} - 6 {\kappa^2 (s)} \bigg(\frac{ds}{dt} \bigg)^2 \frac{d^2s}{dt^2} 
-3{\kappa (s)} {\kappa}^{\prime}(s) \bigg(\frac{ds}{dt} \bigg)^4 = 0.
\end{equation}

Since the function $L$ defined by (\ref{Lagrangian-s}) does not depend explicitly on time, we have the  constant of the motion \cite[Problem 4.10.3]{Kot}
\begin{equation}
\label{Euler-Lagrange-s-constant-motion}
 {\ddot s}\frac{\partial L}{\partial {\ddot s}}  - {\dot s} \bigg[
\frac{d}{dt}\bigg(\frac{\partial L}{\partial {\ddot s}}\bigg)  - \frac{\partial L}{\partial {\dot s}}\bigg] - L = c = \mbox{constant},
\end{equation}
which is a generalization of the Jacobi integral of mechanics \cite{Lemos3}.
By inserting (\ref{Lagrangian-s}) into the above equation we find
\begin{equation}
\label{Euler-Lagrange-s-constant-motion-explicit}
 {\ddot s}^2 - 2\,{\dot s} \, {\dddot s}   + 3\, \kappa^2(s)\, {\dot s}^4 = c.
\end{equation}
Since ${\dot s} (0) =0$, letting $t=0$ in this equation shows that $c$ is  positive.

\subsection{Discomfort induced by the jerk}

We have
\begin{equation}
\label{r-dot-s}
 {\dot {\bf r}} = {\bf r}^{\prime}(s){\dot s} = {\dot s} {\bf t}(s),
\end{equation}
where ${\bf t} = d{\bf r}/ds$  is the unit tangent vector to the curve. By definition 
\begin{equation}
\label{r-primeprime-s}
  \frac{d \bf t}{ds} = {\bf r}^{\prime\prime}(s) = \kappa (s) {\bf n}(s),
\end{equation}
where $\kappa (s) \geq 0$ is the curvature and $\bf n$ is the principal normal vector. Therefore, from equation (\ref{r-dot-s}) it follows that the
acceleration is given by 
\begin{equation}
\label{r-ddot-s}
  {\ddot{\bf r}} = {\ddot s} {\bf t}(s) +  \kappa (s) {\dot s}^2{\bf n}(s).
\end{equation}
Thus the jerk is 
\begin{equation}
\label{jerk-partial}
  {\dddot {\bf r}} = {\dddot s} {\bf t}(s) + {\ddot s}\kappa (s) {\bf n}(s) {\dot s} + {\kappa}^{\prime}(s) {\dot s}^3 {\bf n}(s) + 2 \kappa (s) {\dot s}{\ddot s} {\bf n}(s) + \kappa (s) {\dot s} ^2 \frac{d\bf n}{ds}{\dot s}.
\end{equation}
With the use of  the Frenet (or Frenet-Serret) formula \cite{Manfredo}  
\begin{equation}
\label{Frenet-dot-n}
 \frac{d\bf n}{ds} = - \kappa (s)  {\bf t}(s) + \tau (s) {\bf b}(s),
\end{equation}
where ${\bf b} = {\bf t} \times {\bf n}$ is the binormal vector and $\tau$ is the torsion, we finally obtain
\begin{equation}
\label{r-dddot-jerk}
  {\dddot {\bf r}} = \bigl( {\dddot s} - \kappa^2(s) {\dot s}^3\bigr) {\bf t}(s) + \bigl( 3\kappa (s){\dot s} {\ddot s} + \kappa^{\prime}(s) {\dot s}^3 \bigr) {\bf n}(s) + \kappa (s) \tau (s) {\dot s}^3 {\bf b}(s).
\end{equation}

Since $\{{\bf t}(s),{\bf n}(s),{\bf b}(s) \}$ is an orthonormal set of vectors, we have
\begin{equation}
\label{square-jerk}
\vert  {\dddot {\bf r}} \vert^2 =  \bigl( {\dddot s} - \kappa^2(s) {\dot s}^3\bigr)^2 + \bigl( 3\kappa (s){\dot s} {\ddot s} + \kappa^{\prime}(s) {\dot s}^3 \bigr)^2 + \kappa^2 (s) \tau^2 (s) {\dot s}^6.
\end{equation} 
Thus, the jerk-induced discomfort is given by
\begin{eqnarray}
\label{discomfort-jerk-generalized}
J [s] = \int_0^T \vert  {\dddot {\bf r}} \vert^2 dt & = & \int_0^T \Bigl[  {\dddot s}^2 - 2\kappa^2(s) {\dot s}^3{\dddot s}+
9\kappa^2(s){\dot s}^2 {\ddot s}^2 \nonumber \\
& &  + \,  6\kappa (s)\kappa^{\prime}(s) {\dot s}^4{\ddot s}+  \bigl(\kappa^4 (s) + {\kappa^{\prime}(s)}^2+ \kappa^2(s) \tau^2 (s)\bigr) {\dot s}^6\Bigr] dt.
\end{eqnarray}

The functional (\ref{discomfort-jerk-generalized}) is of the form (\ref{functional-third-order}) with
\begin{equation}
\label{Lagrangian-jerk-discomfort-generalized}
L = {\dddot s}^2 - 2\kappa^2(s) {\dot s}^3{\dddot s}+
9\kappa^2(s){\dot s}^2 {\ddot s}^2   +   6\kappa (s)\kappa^{\prime}(s) {\dot s}^4{\ddot s}+  \bigl(\kappa^4 (s) + {\kappa^{\prime}(s)}^2+ \kappa^2(s) \tau^2 (s)\bigr) {\dot s}^6.
\end{equation}
The boundary conditions for the problem of minimizing the jerk-induced discomfort 
(\ref{discomfort-jerk-generalized}) are
\begin{equation}
\label{boundary-conditions-jerk-generalized}
s(0)=0, \,\,\,\,\,\,\,\,\,\, s(T)=D, \,\,\,\,\,\,\,\,\,\, {\dot s}(0)=0, \,\,\,\,\,\,\,\,\,\, {\dot s}(T)=0.
\end{equation}
Because $\dot s(t)$ vanishes both at $t=0$ and $t=T$, the natural boundary conditions
\begin{equation}
\label{natural-boundary-conditions-jerk-generalized}
\frac{\partial L}{\partial \dddot s}\bigg\vert_{t=0}=0, \,\,\,\,\,\,\,\,\,\, \frac{\partial L}{\partial \dddot s}\bigg\vert_{t=T}=0
\end{equation}
take the same form as in the straight line case, namely
\begin{equation}
\label{natural-boundary-conditions-jerk-generalized-explicit}
{\dddot s}(0)=0, \,\,\,\,\,\,\,\,\,\, {\dddot s}(T)=0.
\end{equation}
The  generalized Euler-Lagrange equation for the functional (\ref{discomfort-jerk-generalized}) --- equation (\ref{Euler-Lagrange-third-order})  with $s$ substituted for $x$ and $L$ given by (\ref{Lagrangian-jerk-discomfort-generalized}) --- is terribly complicated and will not be written down here.

Integration by parts with the use of (\ref{boundary-conditions-jerk-generalized}) 
leads to
\begin{equation}
\label{integration-by-parts-jerk-generalized}
\int_0^T \kappa^2(s){\dot s}^3{\dddot s}dt = \kappa^2(s){\dot s}^3{\ddot s}\bigg\vert_0^T - \int_0^T \frac{d}{dt}\bigl[\kappa^2(s){\dot s}^3\bigr]{\ddot s}dt = - \int_0^T \bigl[2\kappa (s)\kappa^{\prime}(s){\dot s}^4{\ddot s} + 3\kappa^2(s){\dot s}^2{\ddot s}^2\bigr]dt.
\end{equation}
With this result the jerk-induced discomfort (\ref{discomfort-jerk-generalized})
becomes
\begin{equation}
\label{discomfort-jerk-generalized-simplified}
J [s] =  \int_0^T \bigl[  {\dddot s}^2 + 15\kappa^2(s) {\dot s}^2{\ddot s}^2+
10\kappa (s)\kappa^{\prime}(s){\dot s}^4 {\ddot s} 
 +   \bigl(\kappa^4 (s) + {\kappa^{\prime}(s)}^2+ \kappa^2(s) \tau^2 (s)\bigr){\dot s}^6 \bigr] dt.
\end{equation} 
If the   curvature is constant, the discomfort for motion on a curved path is always larger than that for motion on a straight line, which is characterized by  $\kappa = \tau =0$. However, since for variable curvature the third term in the above integrand  may be negative,  the integral of the sum of all terms but the first may conceivably be negative and  one might experience less discomfort by moving on a path with  curvature and torsion than on a straight line, for which the discomfort  would be given by the integral of the first term alone.  Torsion contributes a positive term, but if $\tau =0$ it is not clear  that variable curvature necessarily adds to the jerk-induced discomfort, which appears counterintuitive. This is the reason why the discomfort is seemingly better measured by the acceleration than by the jerk.

\section{Example: circular path}\label{Circular}

Let the path be an arc of a circle of radius $R$ with the starting point of the journey at the origin. 
\begin{figure}[t!]
\epsfxsize=6cm
\begin{center}
\leavevmode
\epsffile{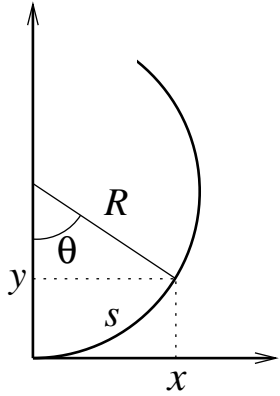}
\caption{Circular arc path.}
\end{center}
\end{figure}
In terms of the angle  $\theta$ shown in Figure 1, the parametric equation of the path is
\begin{equation}
\label{circle-parametric-theta}
 x = R \sin \theta, \,\,\,\,\,\,\,\,\,\, y = R( 1 - \cos \theta ).
\end{equation}
Since $\theta = s/R$, the parameterization of the circle in terms of the arc length is
\begin{equation}
\label{arc-length-parameterized-circle}
{\bf r}(s) = R \sin \Bigl(\frac{s}{R}\Bigr)\, {\hat {\bf x}} + R\Bigl[1- \cos \Bigl( \frac{s}{R}\Bigr)\Bigr] \,{\hat {\bf y}}.
\end{equation}
Thus we have
\begin{equation}
\label{r-primeprime-circle}
{\bf r}^{\prime\prime}(s) = -\frac{1}{R}\Bigl[ \sin \Bigl(\frac{s}{R}\Bigr)\, {\hat {\bf x}} -  \cos \Bigl( \frac{s}{R}\Bigr) \,{\hat {\bf y}}\Bigr], 
\end{equation}
whence
\begin{equation}
\label{r-primeprime-circle}
\kappa^2(s) ={\bf r}^{\prime\prime}(s) \cdot  {\bf r}^{\prime\prime}(s) = \frac{1}{R^2}.
\end{equation}
Consequently, the discomfort to be minimized is 
\begin{equation}
\label{discomfort-functional-acceleration-circle}
A[s] =\int_0^T \Bigl( {\ddot s}^2 + \frac{ {\dot s} ^4}{R^2} \Bigr) dt. 
\end{equation}

We shall circunscribe our discussion to the acceleration-induced discomfort. Since the curvature ${\kappa} = 1/R$ is constant, equation (\ref{Euler-Lagrange-differential-equation-for-s}) takes the simpler  form
\begin{equation}
\label{Euler-Lagrange-differential-equation-for-s-circle}
 \frac{d^4 s}{dt^4} - \frac{6}{R^2} \bigg(\frac{ds}{dt} \bigg)^2 \frac{d^2s}{dt^2}  = 0
\end{equation}
 and its associated constant of the motion (\ref{Euler-Lagrange-s-constant-motion-explicit}) is given by
\begin{equation}
\label{Euler-Lagrange-s-constant-motion-circle}
 {\ddot s}^2 - 2\,{\dot s} \, {\dddot s}   - \frac{3}{R^2}\, {\dot s}^4 = \frac{c}{R^2}
\end{equation}
where the value of the constant of the motion has been written as $c/R^2$ for future convenience. 
As a consistency check, note that in the limit  $R \to \infty$ the circle turns into  a straight  line
and equation (\ref{Euler-Lagrange-differential-equation-for-s-circle}) coincides with equation (\ref{Euler-Lagrange-second-order-acceleration}), as it should. Also, for  very high curvature ($R \to 0$, $\kappa \to \infty$) the discomfort 
(\ref{discomfort-functional-acceleration-circle}) becomes enormous, as intuition suggests.

Our task is to solve the fourth-order nonlinear differential equation  (\ref{Euler-Lagrange-differential-equation-for-s-circle}) with the boundary conditions (\ref{boundary-conditions-s}) and (\ref{boundary-conditions-s-dot}). We start by taking advantage of the constant of the motion  (\ref{Euler-Lagrange-s-constant-motion-circle}), which we write as
\begin{equation}
\label{Euler-Lagrange-s-constant-motion-u-sdot}
 {\dot u}^2 - 2u {\ddot u}   + \frac{3}{R^2}u^4 = \frac{c}{R^2}
\end{equation}
in terms of
\begin{equation}
\label{u-equal-sdot}
 u = {\dot s}.
\end{equation}

The standard trick to lower the order of equation (\ref{Euler-Lagrange-s-constant-motion-u-sdot}) is to introduce a new variable $w$ defined by
\begin{equation}
\label{w-equal-u-dot}
 w = {\dot u}
\end{equation}
and replace the independent variable $t$ by  $u$, so that 
\begin{equation}
\label{ddot-u-interms-w}
 {\ddot u} = \frac{dw}{dt} = \frac{dw}{du}\frac{du}{dt} = w \frac{dw}{du}.
\end{equation}
Thus, equation (\ref{Euler-Lagrange-s-constant-motion-u-sdot}) is reduced to
\begin{equation}
\label{Euler-Lagrange-s-constant-motion-explicit-reduced}
 w^2 - 2u w \frac{dw}{du}  + \frac{3}{R^2}u^4 = \frac{c}{R^2}.
\end{equation}
Setting
\begin{equation}
\label{r-equal-w-squared}
 r = w^2
\end{equation}
we are led to
\begin{equation}
\label{Euler-Lagrange-s-constant-motion-explicit-reduced-r}
 u  \frac{dr}{du}  - r - \frac{3}{R^2}u^4 + \frac{c}{R^2} = 0
\end{equation} 
or, equivalently,
\begin{equation}
\label{Euler-Lagrange-s-constant-motion-explicit-reduced-r-differentials}
 u  dr  +\Bigl(\frac{c}{R^2} - \frac{3}{R^2}u^4 - r \Bigr) du =0.
\end{equation}

Let us attempt to find an integrating factor for the above  differential $1$-form as a function $F$  of $u$ alone.  To this end we write as $\omega = M(u,r)dr+N(u,r)du$ the $1$-form obtained by multiplying the left-hand side of equation (\ref{Euler-Lagrange-s-constant-motion-explicit-reduced-r-differentials}) by $F(u)$. The necessary and sufficient condition for the $1$-form $\omega$   to be exact is $\partial M/\partial u= \partial N/\partial r$. This leads to $uF^{\prime}(u) = -2F(u)$, which is solved by $F(u)=u^{-2}$. Therefore, multiplication of equation (\ref{Euler-Lagrange-s-constant-motion-explicit-reduced-r-differentials})  by  $u^{-2}$ leads to 
\begin{equation}
\label{Euler-Lagrange-s-constant-motion-multiplied-integrating-factor}
 \frac{1}{u}  dr  +\Bigl(\frac{c}{R^2u^2} - \frac{3}{R^2}u^2 - \frac{r}{u^2} \Bigr) du = d\Phi = 0
\end{equation}
where
\begin{equation}
\label{Phi-r-u}
 \Phi (r,u) = \frac{r}{u} - \frac{u^3}{R^2}  - \frac{c}{R^2u}.
\end{equation} 
Thus, by combining this last equation  with (\ref{Euler-Lagrange-s-constant-motion-multiplied-integrating-factor}) we conclude that
\begin{equation}
\label{Phi-r-u-equal-constant}
\frac{r}{u} - \frac{u^3}{R^2}  - \frac{c}{R^2u} = \frac{a}{R^2},
\end{equation} 
where $a$ is a constant.
Taking into account that $r={\dot u}^2$ --- see (\ref{w-equal-u-dot}) and (\ref{r-equal-w-squared})  ---,  equation (\ref{Phi-r-u-equal-constant}) can be rewritten as
\begin{equation}
\label{u-dot-squared}
\Bigl(\frac{du}{dt}\Bigr)^2 = \frac{1}{R^2}(c+ au + u^4).
\end{equation}
Separating variables in the above equation  and integrating we find
\begin{equation}
\label{u-of-t-elliptic-integral}
\int\frac{du}{\sqrt{c+ au  + u^4}} = \frac{t + b}{R},
\end{equation} 
where $b$ is a constant. This integral can be reduced to an elliptic integral, but this is not easy to do explicitly \cite{Lawden}. This means that, by inversion, $u(t)$ will be given by a very complicated combination of elliptic functions. Since $u = {\dot s}$, in order to find $s(t)$ one will still have to compute the integral of this  intricate combination of elliptic functions. This will give $s(t)$ in a quite unmanageable and unilluminating form.

\subsection{Trial functions}

Since the exact solution is unwieldy, we shall first consider analytical  approximate solutions to our minimization problem. Let us first consider the function
\begin{equation}
\label{s-trial-function}
{\tilde s}(t) = D \sin^2\bigg(\frac{\pi t}{2T}\bigg) = \frac{D}{2}\bigg[ 1- \cos \bigg(\frac{\pi t}{T}\bigg) \bigg]. 
\end{equation} 
It follows at once that 
\begin{equation}
\label{s-trial-function-dot}
{\dot {\tilde s}}(t) = \frac{\pi D}{2T}  \sin\bigg(\frac{\pi t}{T}\bigg)
\end{equation} 
and 
\begin{equation}
\label{s-trial-function-dot-dot}
{\ddot {\tilde s}}(t) = \frac{\pi^2 D}{2T^2}  \cos\bigg(\frac{\pi t}{T}\bigg).
\end{equation} 
It is clear from (\ref{s-trial-function}) and (\ref{s-trial-function-dot}) that the boundary conditions (\ref{boundary-conditions-s}) and (\ref{boundary-conditions-s-dot}) are fulfilled. The discomfort brought about  by the trial function (\ref{s-trial-function}) is
\begin{equation}
\label{discomfort-s-trial-function}
A[{\tilde s}] = \int_0^T \bigg[ \frac{\pi^4D^2}{4T^4} \cos^2\bigg( \frac{\pi t}{T}\bigg) + \frac{1}{R^2} \frac{\pi^4D^4}{16T^4} \sin^4\bigg( \frac{\pi t}{T}\bigg)
\bigg] dt.
\end{equation}
With the change of variable $x=\pi t/T$ this simplifies to
\begin{equation}
\label{discomfort-s-trial-function-variable-x}
A[{\tilde s}] = \frac{\pi^3D^2}{4T^3}\int_0^{\pi} \cos^2x dx +  \frac{\pi^3D^4}{16R^2T^3} \int_0^{\pi}\sin^4x dx.
\end{equation}
With the use of 
\begin{equation}
\label{integrals-values}
\int_0^{\pi} \cos^2x dx = \frac{\pi}{2}, \,\,\,\,\,\,\,\,\,\,    \int_0^{\pi}\sin^4x dx = \frac{3\pi}{8}
\end{equation}
we finally have
\begin{equation}
\label{discomfort-s-trial-function-variable-x-final}
A[{\tilde s}] = \frac{\pi^4D^2}{8T^3}\bigg(1+ \frac{3D^2}{16R^2}\bigg).
\end{equation}
This provides an upper bound on the minimum discomfort:
\begin{equation}
\label{discomfort-upper-value}
A_{min} \leq  \frac{\pi^4D^2}{8T^3}\bigg(1+ \frac{3D^2}{16R^2}\bigg).
\end{equation}
As a check, note that in the limit $R \to \infty$ we have $A[{\tilde s}] = \pi^4D^2/{8T^3}$. Since $\pi^4/8 \approx 12.18$, a comparison with (\ref{minimum-A-discomfort}) shows that for motion on a straight line the error committed is only $1.5 \%$, that is, our approximation is pretty good. This suggests that the function (\ref{s-trial-function})  may  furnish a fairly  good approximation to the optimal solution also for finite but not too small $R$ as compared to $D$. 

A better trial function for the straight line case is
\begin{equation}
\label{better-trial-function-straight}
{\bar s} (t) = \frac{D}{2} \bigg[ 1- (1+\alpha ) \cos \Bigl(\frac{\pi t}{T}\Bigr) + \alpha \cos \Bigl(\frac{3\pi t}{T}\Bigr) \bigg]  
\end{equation}
where the real parameter $\alpha$ is to be chosen so as to yield the least discomfort within this class of trial functions. An elementary
calculation yields
\begin{equation}
\label{better-trial-function-straight-discomfort}
A_{str}[{\bar s}] = \int_0^T {\ddot {\bar s}}^2 dt = \frac{\pi^4D^2}{8T^3}\bigl[ (1+ \alpha)^2 + 81 \alpha^2\bigr].
\end{equation}
The minimum of the function of $\alpha$ within brackets is attained at $\alpha = -1/82$. For this value of $\alpha$ we have
\begin{equation}
\label{better-trial-function-straight-discomfort-minimum}
A_{str}[{\bar s}] = 12.027 \frac{D^2}{T^3},
\end{equation}
which is only $0.2 \%$ larger than the true minimum (\ref{minimum-A-discomfort}).

The computation of the discomfort associated with (\ref{better-trial-function-straight})  for the circular path is a bit laborious but requires only the use of  trigonometric identities such as
\begin{equation}
\label{sin3x}
\sin 3x = 3 \sin x - 4 \sin^3 x, \,\,\,\,\,\,\,\,\,\, \cos 3x = 4\cos^3 x - 3\cos x,
\end{equation}
and the following definite integrals, where $n$ is a natural number:
\begin{equation}
\label{useful-integrals}
\int_0^\pi \cos^{2n}x dx  = \int_0^\pi \sin^{2n}x dx = \pi \frac{(2n)!}{(2^n n!)^2}.
\end{equation}
The result for the discomfort (\ref{discomfort-functional-acceleration-circle}) associated with the function (\ref{better-trial-function-straight}) is
\begin{equation}
\label{better-trial-function-discomfort-alpha}
A[{\bar s}]    =  \frac{\pi^4 D^2}{16T^3}\bigg\{ 2(1+ \alpha )^2 + 162 \alpha^2   + \frac{D^2}{R^2} 
\Bigl[ \frac{3}{8} (1+ \alpha )^4 + \frac{3}{2} \alpha (1+ \alpha )^3 + \frac{27}{2} \alpha^2 (1+ \alpha )^2 + \frac{243}{8}\alpha^4\Bigr]\bigg\}.
\end{equation}
Let $f(\alpha )$ be the function  within braces, which  is to be minimized. Setting $f^\prime (\alpha) =0$ one gets a third degree algebraic equation to be solved, whose coefficients depend on the parameter $D/R$. Although a formula in terms of radicals for the roots of a cubic equation is known, it is too complicated to be useful here. Therefore, for a few values of $D/R$ we have numerically computed (for free at the website https://keisan.casio.com) the real root  
of $f^\prime (\alpha) =0$ and calculated the minimum value of $A[{\bar s}]$. In the table below we make a comparison between  $A[{\tilde s}]$, the discomforted generated by the trial function (\ref{s-trial-function}), and $A_{min}[{\bar s}]$, the minimum discomfort brought about by the family of functions (\ref{better-trial-function-straight}), with the values of the discomfort  expressed in units of $D^2/T^3$. 

\begin{center}
\label{table}
\begin{tabular}{|c|c|c|c|}
                    \hline
$D/R$     & Optimal $\alpha$  &  $A_{min}[{\bar s}]$  &  $A[{\tilde s}]$ \\  \cline{1-4}
 1        &     -0.01909      &    14.06              &    14.46     \\ \cline{1-4}
 2        &     -0.03350      &    19.79              &    21.31    \\ \cline{1-4}
 $\pi$    &     -0.04913      &    30.96              &    34.71    \\ \cline{1-4}
 $2\pi$   &     -0.07286      &    76.38              &    102.31    \\
                     \hline  
\end{tabular}
\end{center}
This table shows that the best approximation provided by the family of functions (\ref{better-trial-function-straight}) is systematically better than the approximation furnished by the single function  (\ref{s-trial-function}), and the improvement becomes more and more significant as $D$ increases beyond the semicircumference of the circle.
\begin{figure}[t!]
\epsfxsize=6cm
\begin{center}
\leavevmode
\epsffile{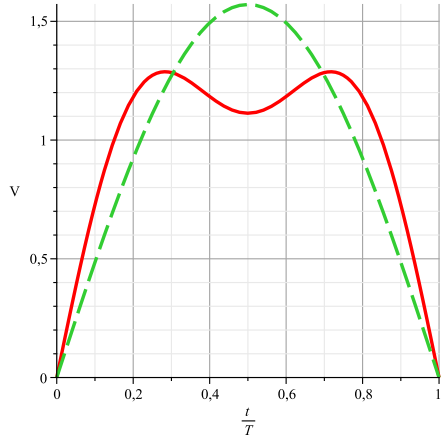}
\caption{ Best speed $ \dot {\bar s}$ (solid line)) within the family (\ref{better-trial-function-straight}) for $\alpha = -0.07286$ (for the case $D/R=2\pi$) compared with $\dot{\tilde s}$ (dashed line) given by (\ref{s-trial-function-dot}). Speeds are  displayed in units of $D/T$.}
\end{center}
\end{figure}
A physical explanation of why, for $D \gg R$, the optimal approximation within the family of functions (\ref{better-trial-function-straight}) is much  better than the approximation afforded by the function (\ref{s-trial-function}) might run as follows. For $D$ big compared to $R$, which would also be the case for fixed $D$ and a highly curved circle,   the discomfort (\ref{discomfort-functional-acceleration-circle}) is dominated by the term involving the speed ${\dot s}$ rather than the one involving the tangential acceleration $\ddot s$. This means that comparatively smaller {\it values} of the speed may result in less discomfort in spite of comparatively larger speed {\it changes}. This is because the pull on the passenger towards the center of the trajectory, connected with the radial (centripetal) acceleration associated with the speed and customarily felt as if being  pressed against the wall of the vehicle, causes more discomfort than speed changes. This is illustrated in Figures 2 and 3 for the case $D=2\pi R$, in which the best approximation within the class (\ref{better-trial-function-straight}) is compared with the approximation (\ref{s-trial-function}). For the better approximation the maximum value attained by the speed is lower, as shown by the solid line in Figure 2, and the smaller contribution to the discomfort of the speed term amply compensates the bigger contribution owing to  the larger speed changes, represented by the solid line in Figure 3.
\begin{figure}[h!]
\epsfxsize=6cm
\begin{center}
\leavevmode
\epsffile{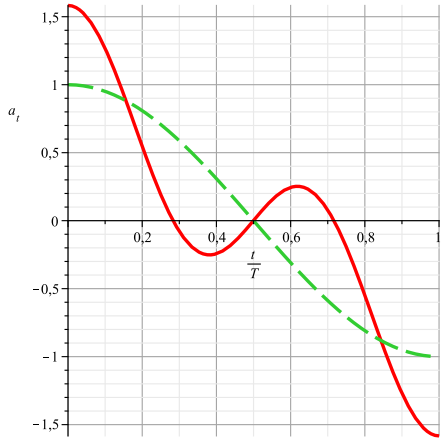}
\caption{Best tangential acceleration  $ \ddot {\bar s}$ (solid line) within the family (\ref{better-trial-function-straight}) for $\alpha = -0.07286$ compared with $\ddot{\tilde s}$ (dashed line) given by (\ref{s-trial-function-dot-dot}). Accelerations are displayed in units of $\pi^2 D/2T^2$.}
\end{center}
\end{figure}

\subsection{Numerical solutions}

In order numerically to solve the ordinary differential equation (\ref{Euler-Lagrange-differential-equation-for-s-circle}) with the boundary conditions (\ref{boundary-conditions-s}) and (\ref{boundary-conditions-s-dot}),  lets us first reformulate the problem in dimensionless form. Let $\tau$ and $\sigma  (\tau )$ be dimensionless quantities defined by
\begin{equation}
\label{s-t-dimensionless}
s(t) = D{\sigma}(\tau ), \,\,\,\,\,\,\,\,\,\, \tau = \frac{t}{T}.
\end{equation}
Note that both $\tau$ and $\sigma (\tau )$ take values in the closed interval $[0,1]$. The chain rule gives
\begin{equation}
\label{s-tilde-derivatives}
\frac{ds}{dt} = D\frac{d\sigma}{d\tau}\frac{d\tau}{dt}= \frac{D}{T}\frac{d\sigma}{d\tau}, \,\,\,\,\,\,\,\,\,\, \frac{d^2s}{dt^2} = \frac{D}{T^2}\frac{d^2\sigma}{d\tau^2},\,\,\,\,\,\,\,\,\,\, \frac{d^4s}{dt^4} = \frac{D}{T^4}\frac{d^4\sigma}{d\tau^4}.
\end{equation}
Substituting these results into (\ref{Euler-Lagrange-differential-equation-for-s-circle}), the dimensionless boundary-value problem to be numerically solved consists of the fourth-order differential equation
\begin{equation}
\label{Euler-Lagrange-differential-equation-for-s-tilde-circle}
 \frac{d^4 \sigma}{d\tau^4} - 6\frac{D^2}{R^2} \bigg(\frac{d\sigma}{d\tau} \bigg)^2 \frac{d^2\sigma}{d\tau^2}  = 0
\end{equation}
together with the boundary conditions
\begin{equation}
\label{boundary-conditions-s-tilde}
 { \sigma}(0)=0, \,\,\,\,\,\,\,\,\,\, { \sigma}(1)=1, \,\,\,\,\,\,\,\,\,\,
{ \sigma}^{\prime}(0)=0, \,\,\,\,\,\,\,\,\,\, 
{ \sigma}^{\prime}(1)=0.
\end{equation}

By means of the Maple software we have numerically solved this boundary value problem for $\sigma (\tau)$, and the figures below compare the speed  for the exact (numerical) solution with the speed that arises from the trial function (\ref{better-trial-function-straight}) for the values of $D/R$ considered in the table above. 





\begin{figure}[!htb]
    \centering
    \begin{minipage}{.5\textwidth}
        \centering
        \includegraphics[width=0.7\linewidth, height=0.3\textheight]{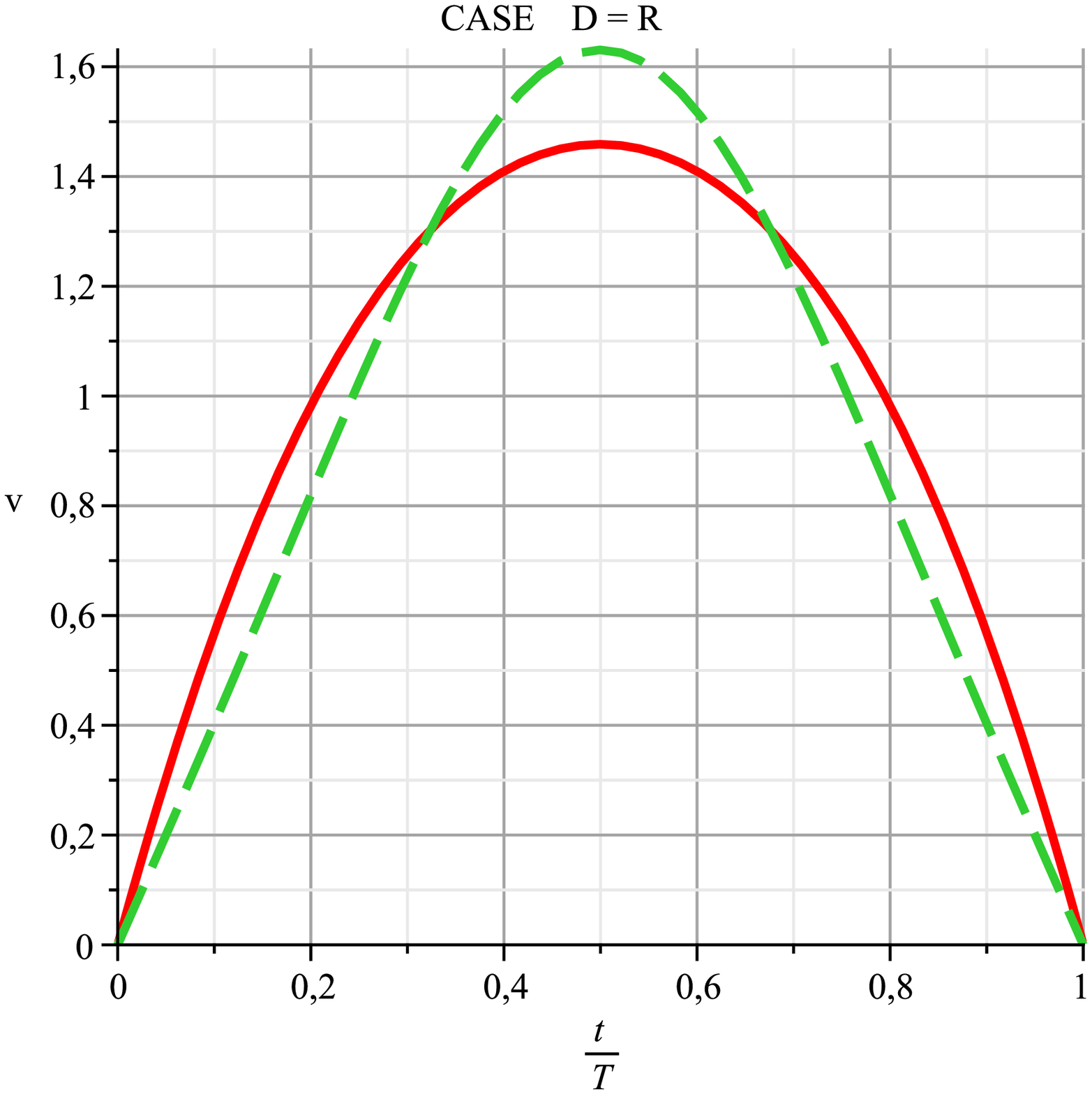}
    \end{minipage}%
    \begin{minipage}{0.5\textwidth}
        \centering
        \includegraphics[width=0.7\linewidth, height=0.3\textheight]{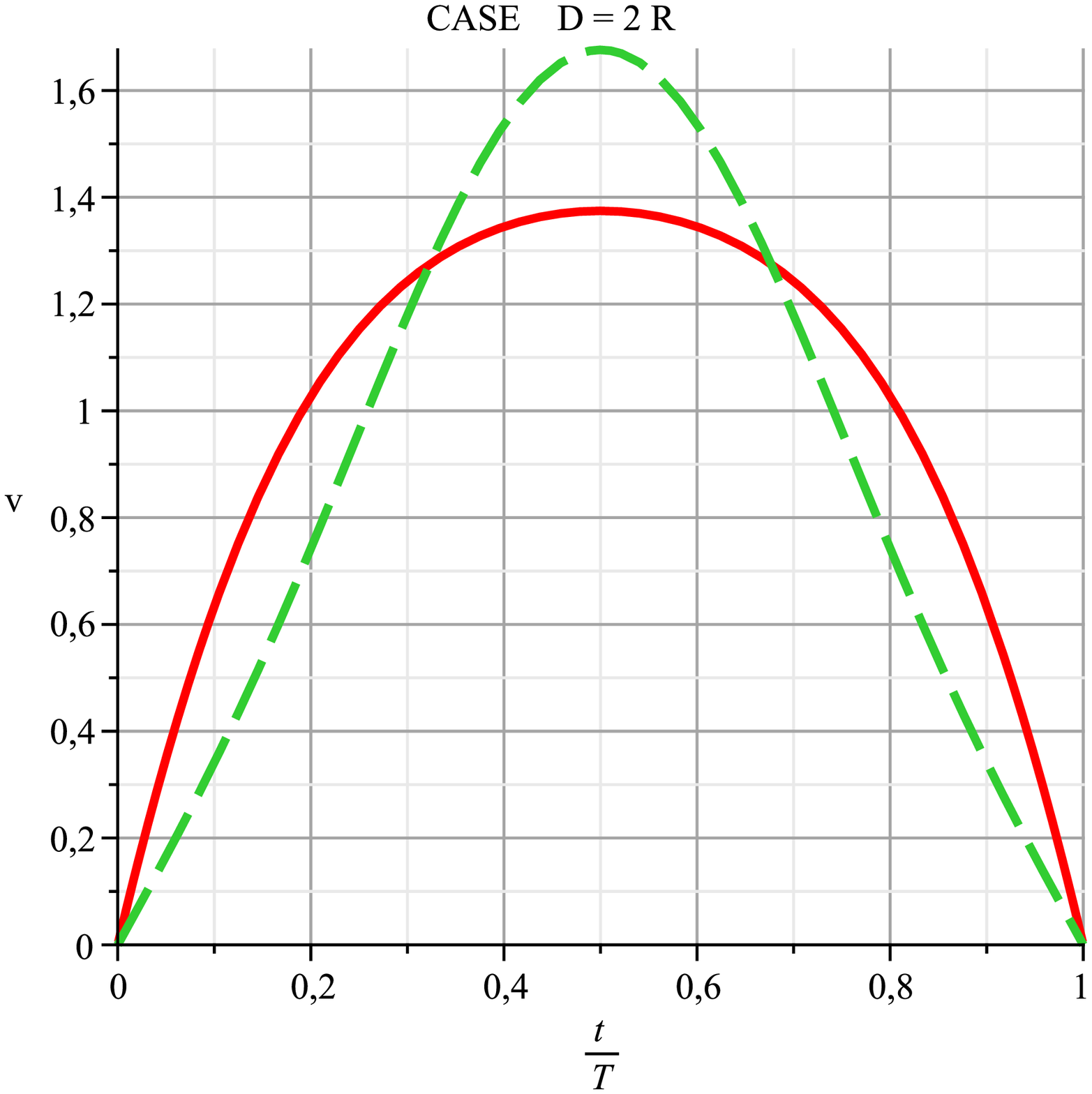}
    \end{minipage}
		\caption{Cases $D=R,\, \alpha = -0.01909$ and $D=2R, \, \alpha = -0.03350$: the numerically computed speed (solid line)  compared with the speed associated with the trial function (\ref{better-trial-function-straight}). Speeds are  displayed in units of $D/T$.}
\label{figura-casos-R-2R}		
\end{figure}

Figure \ref{figura-casos-R-2R} shows that if $D/R \leq 2$ the qualitative behavior of the numerically computed speed is not too different  from that of the straight-line case discussed in \cite{Lemos1}. For $D=R$ the trial  function (\ref{better-trial-function-straight}) provides a tolerable  approximation to the exact solution, but for $D=2R$ the discrepancy is already significant. 

\begin{figure}[!htb]
    \centering
    \begin{minipage}{.5\textwidth}
        \centering
        \includegraphics[width=0.7\linewidth, height=0.3\textheight]{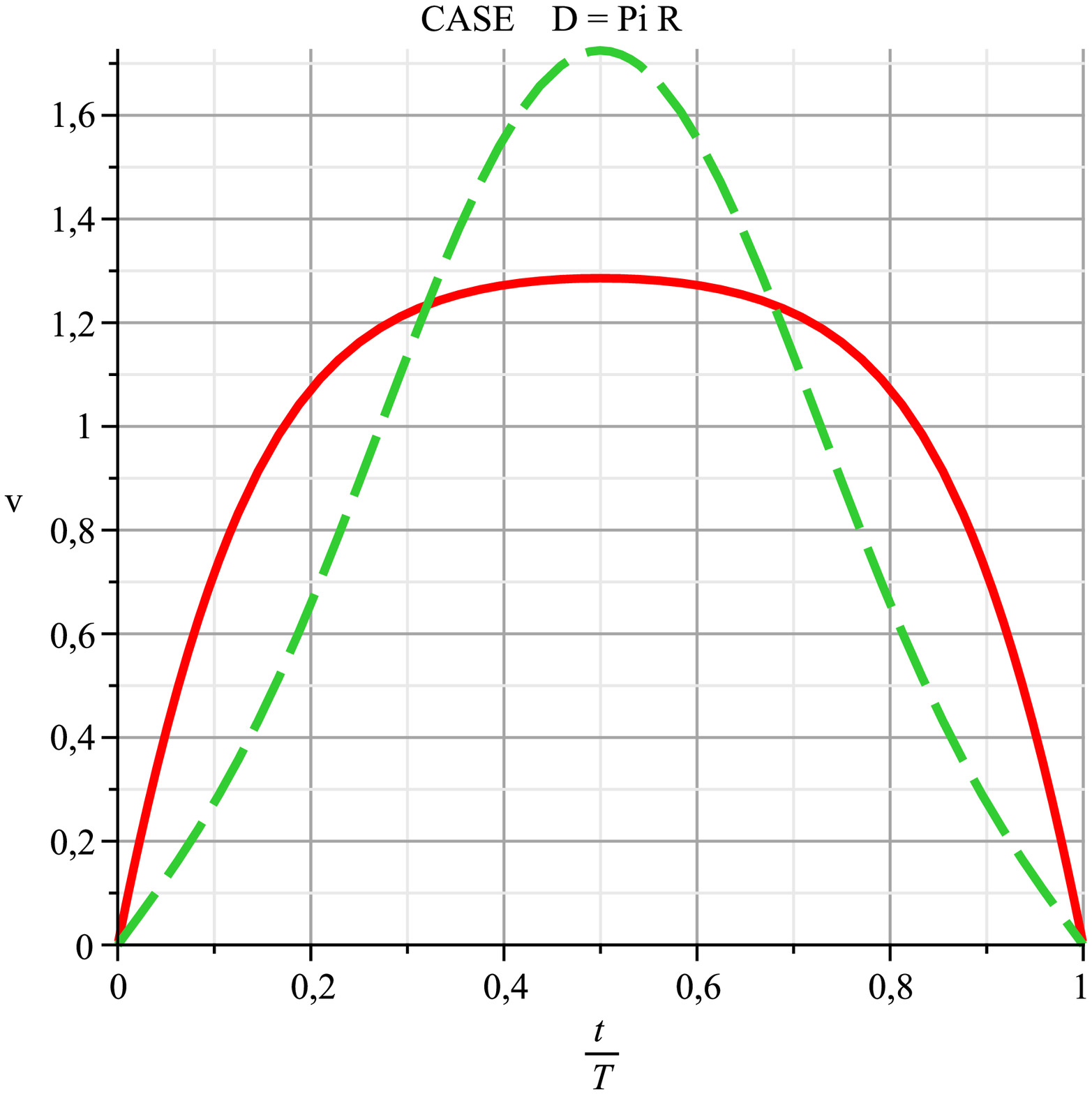}
        \label{fig:prob1_6_2}
    \end{minipage}%
    \begin{minipage}{0.5\textwidth}
        \centering
        \includegraphics[width=0.7\linewidth, height=0.3\textheight]{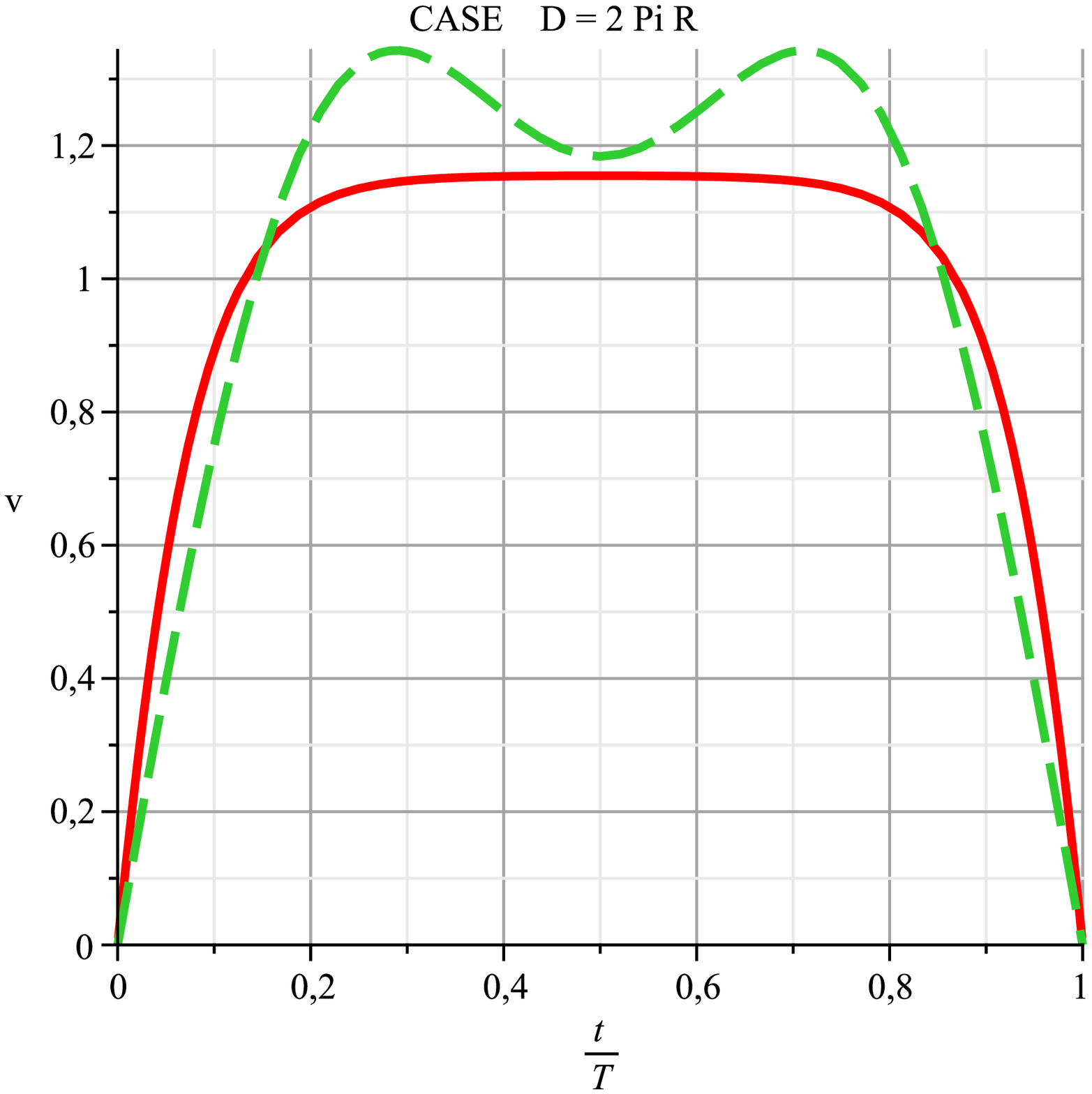}
        \label{fig:prob1_6_1}
    \end{minipage}
		\caption{Cases $D=\pi R,\, \alpha = -0.04913$ and $D=2\pi R, \, \alpha = -0,07286$: the numerically computed speed (solid line)  compared with the speed associated with the trial function (\ref{better-trial-function-straight}). Speeds are  displayed in units of $D/T$.}
\label{figura-casos-PiR-2PiR}
\end{figure}

As can be seen from Figure \ref{figura-casos-PiR-2PiR}, for $D$ equal to or bigger than the semicircumference of the circle the trial function  (\ref{better-trial-function-straight}) is a very bad approximation to the exact solution. As $D$ approaches and  increases beyond the circumference of the circle, 
the strategy to minimize the discomfort becomes clear: accelerate  quickly  from rest to a  speed a little above the average speed $D/T$,  keep this speed  nearly constant for most of the journey, then slow down quickly to a full stop. 

\begin{figure}[h!]
\epsfxsize=6cm
\begin{center}
\leavevmode
\epsffile{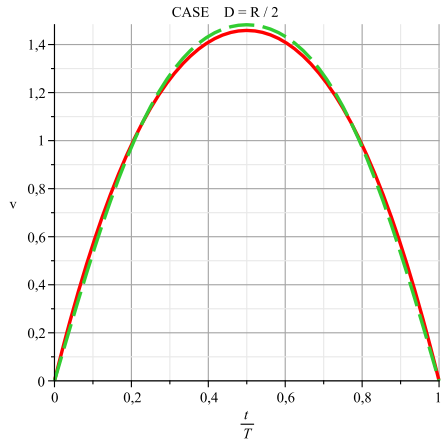}
\caption{Case $D=R/2,\, \alpha = -0.01406$: the numerically computed speed (solid line)  compared with the speed associated with the trial function (\ref{better-trial-function-straight}). Speeds are  displayed in units of $D/T$.}
\label{figura-caso-2D-R}
\end{center}
\end{figure}

The trial  function (\ref{better-trial-function-straight}) provides a very poor approximation to the exact solution for $D/R \sim 2$ and becomes worse and worse as the ratio $D/R$ increases beyond two. An improvement would require a trial function depending on  more than one parameter. But this would make it extremely cumbersome to  compute the integral (\ref{discomfort-functional-acceleration-circle}) and  subsequently solve the problem of minimizing a function of several variables.  Let us note, nevertheless, that the trial function (\ref{better-trial-function-straight}) provides an excellent approximation to the exact solution for $D \leq R/2$, as Figure \ref{figura-caso-2D-R} illustrates very well. 

\section{Final Remarks}\label{Final-remarks}

The problem of the least uncomfortable journey  affords the opportunity to acquaint advanced undergraduate or beginning graduate students with several useful techniques of the calculus of variations. It illustrates nicely that the appropriate choice of independent and dependent variables can make a variational problem much more tractable, sometimes allowing us to get rid of constraints.  It   also provides a nice example of a higher-order variational problem of physical interest.  

The generalization of the  least uncomfortable journey problem to motion on an arbitrary path makes it possible to compare the measures of discomfort induced by acceleration and jerk. The acceleration-induced  discomfort is always larger for a curved path than for a straight line,  as intuitively expected. This is not necessarily so for the jerk-induced  discomfort, which suggests that  the integral of the square of the acceleration  provides a more reliable measure   
of the discomfort than the one supplied by the jerk.

\subsection*{Acknowledgment}
The author is thankful to the anonymous reviewer whose suggestions helped to improve the manuscript significantly.


 
 \bibliographystyle{unsrt}

\end{document}